Hydrostatic pressure study of pure and doped La$_{1-x}$R$_x$AgSb$_2$ (R = Ce, Nd) charge-density-wave compounds

**M. S. Torikachvili**

*Department of Physics, San Diego State University, San Diego, CA 92182-1233 and Ames Laboratory, Iowa State University, Ames, IA 50011, USA*

**S. L. Bud'ko, S. A. Law,\* M. E. Tillman, E. D. Mun, and P. C. Canfield**

*Ames Laboratory and Department of Physics and Astronomy, Iowa State University, Ames, IA 50011, USA*

Abstract

The intermetallic compound LaAgSb$_2$ displays two charge-density-wave (CDW) transitions, which were detected with measurements of electrical resistivity ($\rho$), magnetic susceptibility, and X-ray scattering; the upper transition takes place at $T_1 \approx 210$ K, and it is accompanied by a large anomaly in $\rho(T)$, whereas the lower transition is marked by a much more subtle anomaly at $T_2 \approx 185$ K. We studied the effect of hydrostatic pressure *(P)* on the formation of the upper CDW state in pure and doped La$_{1-x}$R$_x$AgSb$_2$ *(R = Ce, Nd)* compounds, by means of measurements of $\rho(T)$ for *P ≤ 23 kbar*. We found that the hydrostatic pressure, as well as the chemical pressure introduced by the partial substitution of the smaller Ce and Nd ions for La, result in the suppression of the CDW ground state, e.g. the reduction of the ordering temperature $T_1$. The values of $dT_1/dP$ are $\approx$ 2-4 times higher for the Ce-doped samples as compared to pure LaAgSb$_2$, or even La$_{0.75}$Nd$_{0.25}$AgSb$_2$ Nd-doped with a comparable $T_1$ *(P=0)*. This increased sensitivity to pressure may be due to increasing Ce–hybridization under pressure. The magnetic ordering temperature of the cerium-doped compounds is also reduced by pressure, and the high pressure behavior of the Ce-doped samples is dominated by Kondo impurity scattering.

## 1. Introduction

The intermetallic ternary compounds $R$AgSb$_2$ ($R$ = rare earth) have remarkable electronic and magnetic properties. These materials crystallize in a simple tetragonal ZrCuSi$_2$-type structure (P4/nmm, No. 129),[1,2] where planar layers of Sb, Ag, and $R$-Sb are stacked along the c-axis.[3] This crystallographic anisotropy manifests itself in moderately high resistivity anisotropy ($\rho_c$ is approximately ten times larger than $\rho_{ab}$), as well as in Fermi surface nesting, and two charge-density wave (CDW) transitions in LaAgSb$_2$.[3,4]

Single-crystals for most of the $R$AgSb$_2$ have been grown from a Sb-rich flux, with the exceptions of $R$ = Pm, Eu, Yb, and Lu.[3] The $\rho_{ab}(T)$, and magnetic susceptibility data for LaAgSb$_2$ show a pronounced features at $T_1 \approx 210$ K,[3] and a much more subtle anomaly at $T_2 \approx 185$ K,[4] both of which are reminiscent of CDW transitions. An X-ray scattering study confirmed that the origin of these 2 features in $\rho(T)$ is the formation of CDWs.[4] Below $T_1$ a CDW modulation with wave vector $\approx$ (0.026, 0, 0) develops along the a- and b- directions, whereas the CDW below $T_2$ develops along the c-direction, with a wave vector of (0, 0, 0.166).[4] A hydrostatic pressure study up to 8 kbar showed that the onset of the CDW at $T_1$ was lowered at the rate $dT_1/dP \approx -4.3$ K/kbar.[5]

The local moment bearing $R$AgSb$_2$ compounds display a rich variety of magnetic phenomena, including long range magnetic order at low temperatures with Néel temperatures ($T_N$) ranging from 2 – 11 K.[1,3] Neutron diffraction data show that the $R$AgSb$_2$ ($R$ = Pr, Nd, Tb, Dy, Ho, Er, Tm) compounds exhibit collinear antiferromagnetic order.[6] The value of $T_N$ for the heavy $R$ scales relatively well with the de Gennes factor $(g_J-1)^2(J(J+1))$ suggesting that the long-range ordering is mediated by RKKY indirect exchange.[3,6] On the other hand CeAgSb$_2$ orders ferromagnetically with a Curie temperature ($T_m$) $\approx$ 9.5 K.[3,6] In light of the Ce-Ce ion distance being 4.39 Å, which exceeds the Hill limit by about 30%,[7] magnetic ordering (rather than full, Kondo screening), is indeed expected. The effective moment of the Ce-ions inferred from a Curie-Weiss analysis of the temperature dependent magnetic susceptibility is 2.26 $\mu_B$,[3] somewhat reduced from the Ce$^{3+}$ value of 2.54 $\mu_B$; whereas low temperature magnetization isotherms, and the neutron diffraction work indicated that the Ce-ion magnetic moments are aligned along the c-axis, and that the magnitude of the ordered moment is 0.33 $\mu_B$.[3,6] The temperature dependence of the electrical resistivity of CeAgSb$_2$ in the paramagnetic phase is typical of a Kondo lattice system: the value of $\rho$ drops slightly below 300 K, it passes through a shallow minimum near 175 K, and it starts to rise again at lower temperatures, at a faster rate. $\rho(T)$ passes through a maximum near 18 K, and then it drops sharply below $T_m \approx 9.7$ K, consistently with the suppression of spin-flip scattering below $T_m$.[3] The value of $T_m \approx 9.7$ K for CeAgSb$_2$ is gradually depressed by dilution with La, reaching 3.2 K for La$_{0.8}$Ce$_{0.2}$AgSb$_2$.[5, 8]

Similarly to the effect of pressure on the CDW order in LaAgSb$_2$, the partial substitution of the smaller Y, Ce, Nd, and Gd for La in the La$_{1-x}$R$_x$AgSb$_2$ compounds lowers the value of $T_1$, and the CDW transition appears to be completely suppressed for $x \approx 0.3$.[5, 8] The magnitude of the depression of $T_1$ both due to pressure, or as a result of partial substitution of Y, Ce, Nd, and Gd for La scales with the c/a ratio, the latter at a much faster rate, possibly highlighting an additional detrimental effect of the disorder to the stability of the CDW.[5]

In order to further probe the stability of the CDW ground state in La$_{1-x}$R$_x$AgSb$_2$, and to try to further separate and clarify the effects of hydrostatic pressure, chemical pressure, and disorder, we carried out measurements of in-plane $\rho(T)$ of pure, and doped La$_{1-x}$R$_x$AgSb$_2$ materials, where $R$ = Ce ($x$ = 0.1, 0.2, and 0.3), and Nd ($x$ = 0.25), for hydrostatic pressures up to 23 kbar. In light of its pronounced feature in $\rho(T)$ we focused on the higher temperature CDW transition ($T_1$, subsequently referred to as just $T_{CDW}$), and we monitored its dependence both as a function of composition and pressure.

## 2. Experimental Details

Single-crystals of the La$_{1-x}$R$_x$AgSb$_2$ materials $R$ = Ce, and Nd were grown from Sb-rich solutions.[3, 9] The crystals form with a plate-like habit; the typical dimensions are 5 x 5 x 2 mm$^3$, and the large plate surface is perpendicular to the crystallographic c-axis. The crystals used for the $\rho(T, P)$ measurements were cut and polished into a bar shape with typical dimensions of 3 x 1 x 0.5 mm$^3$, with the longer dimension, the direction of the current flow, being close to the (100) direction. Four platinum leads were attached to the samples with Epotek H20E Ag-loaded epoxy. Ambient pressure measurements of $\rho(T)$ were carried out with a Quantum Design Physical Property Measurement System (PPMS-9), using a low-frequency ac technique. Measurements of $\rho(T, P)$ for hydrostatic pressures up to $\approx$ 23 kbar were carried out using a self-clamping piston-cylinder Be-Cu pressure cell, with a hardened NiCrAl (40HNU-VI) core. For the high-pressure measurements, the sample, a coil of manganin, and a coil of Sn were connected to the 12 wires at the end of a Stycast-sealed feed through. This assembly was inserted into a Teflon-cup filled with a 40:60 mixture of mineral oil:n-pentane, which served as the pressure transmitting medium. Four-wire measurements of the resistance $R$ of sample, and pseudo-four-wire measurements of the manganin and Sn were carried out using a Linear Research LR-700 ac resistance bridge, operating at 16 Hz. Pressure was applied and locked in the cell at ambient temperature using a hydraulic press. The pressure at room temperature was determined from the change in resistance of the manganin manometer, and at low $T$ from the superconducting transition temperature of Sn.[10] The internal pressure in this type of cell is reduced upon cooling,[11] reflecting the different thermal expansion characteristics of the cell body and the pressure-transmitting medium. In order to estimate the value of the pressure between the two end-range temperatures where the pressure could be determined accurately, $\approx$ 300 K, and $\approx$ 4 K, we assumed a linear reduction of $P$ between 300 K and 90 K,[11] and neglected any $P$ changes between 2 and 90 K. These $P$ values coincided, within a few percent, with the values estimated from the $R(T)$ measurements of the manganin manometer. The pressure cell was mounted and thermally anchored to the sample holder in a flow cryostat, and the temperature was monitored with a Si-diode placed on the sample holder. Resistance measurements were taken while the temperature was varied slowly. In order to minimize the lag in temperature between the sample and temperature sensor, the rate of change of temperature was kept below 0.5 K/min.

## 3. Experimental Results

*LaAgSb$_2$*

The effect of hydrostatic pressure on the behavior of $\rho(T)$ is shown in Fig. 1. The ambient pressure residual-resistivity-ratio for this sample is close to 90, which attests for the quality of the single-crystal, and permits an analysis of the effect of pressure on the CDW order unencumbered by the presence of defects. These $\rho(T)$ data show that pressure lowers the onset temperature for the CDW ordered state at the rate $dT_{CDW}/dP \approx$ -4.3 K/kbar over the whole pressure range studied, as shown in Fig. 2. The $\rho(T)$ data in Fig. 1 also clearly show that the resistivity of LaAgSb$_2$ is reduced by pressure in a similar manner both above and below T$_{CDW}$. This can be quantified by examining $\Delta\rho/\rho = (\rho(P,T)-\rho(0,T))/\rho(0,T)$ *versus P*, for two temperatures: one well below and the other well above $T_{CDW}$. Fig. 3 shows that the value of $\Delta\rho/\rho$ decreases with pressure at approximately the same rate $d(\rho/\rho_0)/dP \approx -8.8 \times 10^{-3}$ kbar$^{-1}$, suggesting that the density of states at the Fermi level increases with pressure, and that the partial energy gap formed at $T_{CDW}$ involves a small portion of the Fermi surface.

The extent to which the CDW gaps the Fermi surface and the effect that pressure has on the gap formation can be quantified by evaluating the size of the increase in $\rho$, just below $T_{CDW}$. This increase is most clearly seen in the difference between the $\rho(T)$ data and the extrapolation of the $T > T_{CDW}$ data to just below $T_{CDW}$. The $\Delta\rho(T)$ data for the various pressures can be compared to each other by normalizing each data set to the resistivity just before the CDW transition. These data are plotted on an effective temperature ($T/T_{CDW}$) scale in Fig. 4. The resistive anomaly associated with the CDW formation is remarkably robust over the whole pressure range, remaining sharp and clearly discernible. A simple caliper of how much Fermi surface is gapped is to take the maximum value of $\Delta\rho(T)/\rho(T_{CDW})$ data for each pressure. These data are plotted in Fig. 2 along with the pressure dependence of $T_{CDW}$. The degree of Fermi surface nesting appears to be suppressed by pressure at a rate that is comparable or slightly slower than $T_{CDW}$, and a linear extrapolation of $T_{CDW}$ to zero provides an upper estimate of P $\approx$ 50 kbar to fully suppress the CDW ground state.

*$La_{0.75}Nd_{0.25}AgSb_2$*

The partial substitution of Nd for La leads to a depression in $T_{CDW}$ at a rate slightly lower than -4 K/at%Nd. $La_{0.75}Nd_{0.25}AgSb_2$, at atmospheric pressure, transforms into a low-temperature CDW state at nearly the same temperature as pure $LaAgSb_2$ when it is under 21.2 kbar pressure (Fig. 5). Although the transition is broadened by the addition of Nd, possibly due to the introduction of disorder, the size of the resistive anomaly (and therefore the degree of Fermi surface nesting) found in $La_{0.75}Nd_{0.25}AgSb_2$ is comparable to that seen in $LaAgSb_2$ at 21.2 kbar.

The $\rho(T)$ curves for $La_{0.75}Nd_{0.25}AgSb_2$ under pressures up to 22 kbar are shown in Fig. 6. The effect of pressure is to further suppress $T_{CDW}$ and the magnitude of the resistive anomaly, until it can no longer be clearly distinguished for $P$ higher than $\approx$ 10 kbar. In order to follow the suppression of $T_{CDW}$ as far as possible, we analyzed the $d\rho/dT$ derivatives taken from the $\rho(T)$ data shown in Fig. 6. For P < 10 kbar $T_{CDW}$ is identified via the local minima in d$\rho$/dT, as shown in Fig. 7. For $P > 10$ kbar this feature is no longer clearly identifiable. The initial pressure dependence of $T_{CDW}$ in $La_{0.75}Nd_{0.25}AgSb_2$ is

about - 5.7 K/kbar, which is not too different from the -4.3 K/kbar value found for LaAgSb$_2$.

$La_{1-x}Ce_xAgSb_2$ (x = 0.1, 0.2, 0.3)

The temperature dependence of the electrical resistivity data for pressures up to $\approx$ 23 kbar for the La$_{1-x}$Ce$_x$AgSb$_2$ ($x$ = 0.1, 0.2, 0.3) compounds is shown in Figs. 8 and 9. These data are shown as curves of normalized $\rho/\rho_{300K}$ vs $T$, and the data for the various pressures are offset for clarity. The partial substitution of Ce for La drives $T_{CDW}$ down at the rate of $\approx$ –5 K/at%Ce, until the CDW feature in $\rho(T)$ can no longer be identified in the La$_{0.7}$Ce$_{0.3}$AgSb$_2$ material.

The effect of pressure on La$_{0.9}$Ce$_{0.1}$AgSb$_2$ is show in Fig. 8. Even at ambient pressure the resistive anomaly associated with the CDW transition is considerably broader than that found for either pure LaAgSb$_2$ or La$_{0.75}$Nd$_{0.25}$AgSb$_2$, underscoring the effect of Kondo scattering, in addition to the disorder introduced by the partial substitution of Ce for La. Examination of Fig. 8a reveals that the effects of pressure appear to be 1) a suppression of $T_{CDW}$, 2) a further distortion of the resistive anomaly associated with the CDW state, and 3) at higher pressures, the complete emergence of what appears to be a single ion Kondo impurity scattering: a resistance minimum (at $T_{min}$) followed by a lower temperature upturn. For lower pressure values the CDW anomaly and the growing Kondo scattering effects combine to produce the apparent distortion of the CDW anomaly. This can be seen in Figs. 8b and 8c where the P = 12.9 kbar data (taken to be representative of the Kondo scattering with $T_{CDW}$ fully suppressed) is subtracted from the $P$ = 7.8 and 10.3 kbar data, respectively, both of which still manifest a CDW anomaly. In both cases the CDW resistive anomalies, which are qualitatively similar to that seen at ambient pressure, are found to be superimposed on the Kondo scattering contribution. In La$_{0.9}$Ce$_{0.1}$AgSb$_2$. $T_{CDW}$ is suppressed at a much higher rate ($\approx$ -9 K/kbar) than in either LaAgSb$_2$ or La$_{0.75}$Nd$_{0.25}$AgSb$_2$.

The $\rho(T)$ curves for La$_{0.8}$Ce$_{0.2}$AgSb$_2$ and La$_{0.7}$Ce$_{0.3}$AgSb$_2$ under pressures up to 22 kbar are shown in Fig. 9. The data for La$_{0.8}$Ce$_{0.2}$AgSb$_2$ are qualitatively similar to those for La$_{0.9}$Ce$_{0.1}$AgSb$_2$ where, at lower pressures, the resistive anomaly from the CDW appears superimposed on the Kondo scattering contribution, leading to a Kondo groundstate when pressure drives $T_{CDW}$ below the temperature of the Kondo-driven minimum in $\rho(T)$. For La$_{0.8}$Ce$_{0.2}$AgSb$_2$, $T_{CDW}$ is suppressed by pressure even more rapidly than La$_{0.9}$Ce$_{0.1}$AgSb$_2$, at a rate of $\approx$ -16 K/kbar. For $La_{0.7}Ce_{0.3}AgSb_2$ the CDW transition appears to be completely suppressed. Both $La_{0.8}Ce_{0.2}AgSb_2$ and $La_{0.7}Ce_{0.3}AgSb_2$ show clear evidence of ferromagnetic order at low temperatures,[5] which manifests itself in the resistivity data as a sharp loss of spin-disorder scattering. This feature is rapidly suppressed by the application of pressure. A tentative phase diagram displaying the range of the CDW, Kondo, and ferromagnetic groundstates as a function of Ce content, temperature, and pressure is shown in Fig. 10. It is worth noting that whereas $dT_{CDW}/dP$ varies by almost a factor of two between the $La_{0.9}Ce_{0.1}AgSb_2$ and $La_{0.8}Ce_{0.2}AgSb_2$ samples, $dT_m/dP \approx$ -0.2 K/kbar is almost the same for $x = 0.2$ and 0.3 (and similar to the initial pressure derivative for pure $CeAgSb_2$),[12] within our limited range of detection.

**4. Discussion and Summary**

The CDW transition in $LaAgSb_2$ is remarkably robust and tunable: modest hydrostatic pressures ($P < 25$ kbar) can lower $T_{CDW}$ by almost 50% and it appears likely that the CDW anomaly will remain sharp and well defined for even larger pressures and suppressions. The partial substitution of Nd or Ce for La introduces lattice disorder, leading to a less stable CDW state, the suppression of $T_{CDW}$, and the broadening of the anomalies in $\rho(T)$ at $T_{CDW}$.[5] The further application of hydrostatic pressure to these doped samples drives the CDW transition even closer to zero temperature.

Figure 11 plots the pressure dependence of $T_{CDW}$ for $LaAgSb_2$, $La_{0.75}Nd_{0.25}AgSb_2$, and $La_{1-x}Ce_xAgSb_2$ ($x = 0.1, 0.2$). The pressure dependence of $T_{CDW}$ in $LaAgSb_2$ and $La_{0.75}Nd_{0.25}AgSb_2$ are similar and the addition of Nd can be rationalized as producing an internal, chemical, pressure. This can be seen graphically in Fig. 11 by shifting the $T_{CDW}$ data for $La_{0.75}Nd_{0.25}AgSb_2$ by 22.4 kbar. On the other hand the pressure dependence of $T_{CDW}$ for $La_{0.9}Ce_{0.1}AgSb_2$ and $La_{0.8}Ce_{0.2}AgSb_2$ are significantly larger than those of $LaAgSb_2$ and $La_{0.75}Nd_{0.25}AgSb_2$. Although the ambient pressure suppression of the Ce substituted compounds appears to be consistent with the chemical pressure model, the enhanced pressure dependence indicates that other effects need to be taken into account.

Figs. 8 and 9 clearly demonstrate that the Kondo effect plays a key role in the low temperature properties of the $La_{1-x}Ce_xAgSb_2$ compounds. One way of understanding the enhanced suppression of $T_{CDW}$ in the Ce-substituted materials is to invoke the possibility that the enhanced pressure sensitivity is associated with the hybridization of the Ce-ions. This proposed, additional, pressure sensitivity associated with Ce is consistent with the fact that the enhanced $dT_{CDW}/dP$ appears to scale with the degree of Ce content, approximately doubling with the change from $x = 0.1$ to $x = 0.2$.

Although the effects of the Kondo impurity scattering and CDW formation appear to be essentially additive when $T_{CDW}$ is larger than the temperature of the Kondo, resistive minimum, it is worth noting that the CDW transition seems to suddenly disappear when pressure drives $T_{CDW}$ down towards $T_{min}$, as indicated in Fig. 10. This brings up the intriguing question of whether the Fermi surface nesting, which is so vital for the CDW formation, can take place once the Kondo ground state is forming. In addition, if the Kondo screening is somehow detrimental to the CDW ground state, can re-entrant behavior be observed? The careful choice of Ce content, $x$, as well as fine control of $P$, may well allow the $La_{1-x}Ce_xAgSb_2$ system to address these questions.

*Acknowledgments*

The authors would like to thank J.C. Fredricks for his help with sample preparation. The support from NSF Grant No. DMR-0306165 for the work at SDSU, and from the US DOE Contract DE-AC02-07CH11358 for work at Ames Laboratory/ISU are gratefully acknowledged.

REFERENCES

* current address – Department of Physics, University of Illinois at Urbana-Champaign, Urbana, IL, 61801-3080

Figure Captions

Figure 1. (Color online) $\rho(T)$ curves for LaAgSb$_2$ in the hydrostatic pressures $P_{300K} = 0$, 2.5, 7.0, 11.1, 15.0, 19.4, and 21.2 kbar. The onset of CDW order is marked by the break in the linear behavior of $\rho(T)$ at high temperatures. Note that the value of $P$ is reduced somewhat upon cooling.

Figure 2. (Color online) Pressure dependence of $T_{CDW}$ (solid circles), and $\Delta\rho_{max}/\rho_{CDW}$ (open squares) for LaAgSb$_2$. The $P$ values shown are corrected to account for the $P$ reduction upon cooling, and they represent the $P$ values at $T_{CDW}$. The values of $\Delta\rho$ for $\Delta\rho_{max}/\rho_{CDW}$ are extracted from the maximum differences between $\rho$ for $T < T_{CDW}$, and the linearly extrapolated values of $\rho$ from $T > T_{CDW}$ to $T < T_{CDW}$.

Figure 3. (Color online) Pressure dependence of $\Delta\rho/\rho$ (see text) at 50 and 300 K, well below and well above $T_{CDW}$, respectively. The $P$ values at 50 K have been corrected to reflect the reduction in pressure with temperature. The lines are from linear fits to the data.

Figure 4. (Color online) Normalized change in resistivity below $T_{CDW}$ as a function of effective temperature (see text). These data were extracted from the data shown in Fig. 1. The pressure values shown are corrected to reflect the estimated values at $T_{CDW}$. For clarity, only a small subset of the data is shown.

Figure 5. (Color online) Normalized curves of $\rho/\rho_{300K}$ vs $T$ for La$_{0.75}$Nd$_{0.25}$AgSb$_2$ and LaAgSb$_2$ at $P_{300K} = 0$, and 21.2 kbar, respectively. The similar values of $T_{CDW}$ and resistive anomalies below $T_{CDW}$ suggest similar degrees of nesting of the Fermi surface.

Figure 6. (Color online) Normalized electrical resistivity $\rho/\rho_{300K}$ vs $T$ for La$_{0.75}$Nd$_{0.25}$AgSb$_2$ compounds in pressures up to 22 kbar. The $\rho/\rho_{300K}(T)$ curves for $P > 0$ are offset for clarity.

Figure 7. (Color online) $d\rho/dT$ as a function of temperature for La$_{0.75}$Nd$_{0.25}$AgSb$_2$ at differing applied pressures. Pressure values shown have been corrected for differential thermal contraction and represent estimated values at $T_{CDW}$. The $\rho(T)$ data have been smoothed before the derivatives were taken. The highest pressure in which $T_{CDW}$ can be unequivocally identified is 9.6 kbar.

Figure 8. (Color online) (a) Normalized electrical resistivity $\rho/\rho_{300K}$ vs $T$ for the La$_{0.9}$Ce$_{0.1}$AgSb$_2$ under pressures up to 22 kbar. The $\rho/\rho_{300K}(T)$ curves for $P > 0$ are offset for clarity; (b) and (c) subtraction of $\rho_{12.9kbar}(T)$ (suppressed CDW) from $\rho_{7.8kbar}(T)$, and $\rho_{10.3kbar}(T)$ ($T_{CDW}$ = 116 K, and 90 K, respectively), yielding $\rho_{CDW}(T)$ behavior deconvoluted from the contribution of single-ion Kondo scattering.

Figure 9. (Color online) Normalized electrical resistivity $\rho/\rho_{300K}$ vs $T$ for (a) La$_{0.9}$Ce$_{0.2}$AgSb$_2$, and (b) La$_{0.7}$Ce$_{0.3}$AgSb$_2$ in pressures up to 22 kbar. The $\rho/\rho_{300K}(T)$ curves for $P > 0$ are offset for clarity.

Figure 10. (Color online) Pressure dependence of $T_{CDW}$, $T_{min}$, and $T_m$ for La$_{1-x}$Ce$_x$AgSb$_2$. The inset shows the detailed values of $T_m(P)$ for $x$ = 0.2, and 0.3. The value of $dT_m/dP$ for both compositions is $\approx$ -0.2 K/kbar.

Figure 11. (Color online) Temperature dependence of $T_{CDW}$ as a function of pressure for LaAgSb$_2$, La$_{1-x}$Ce$_x$AgSb$_2$ ($x$ = 0.1, 0.2), and La$_{0.75}$Nd$_{0.25}$AgSb$_2$. The open symbols for the doped materials represent actual $T_{CDW}$s. The solid symbols for the doped materials were shifted in pressure so that their ambient pressure values match the $T_{CDW}$ values for pure LaAgSb$_2$, in order to emphasize the stronger detrimental effect of pressure on the CDW ground state for the doped materials, particularly for Ce doping.

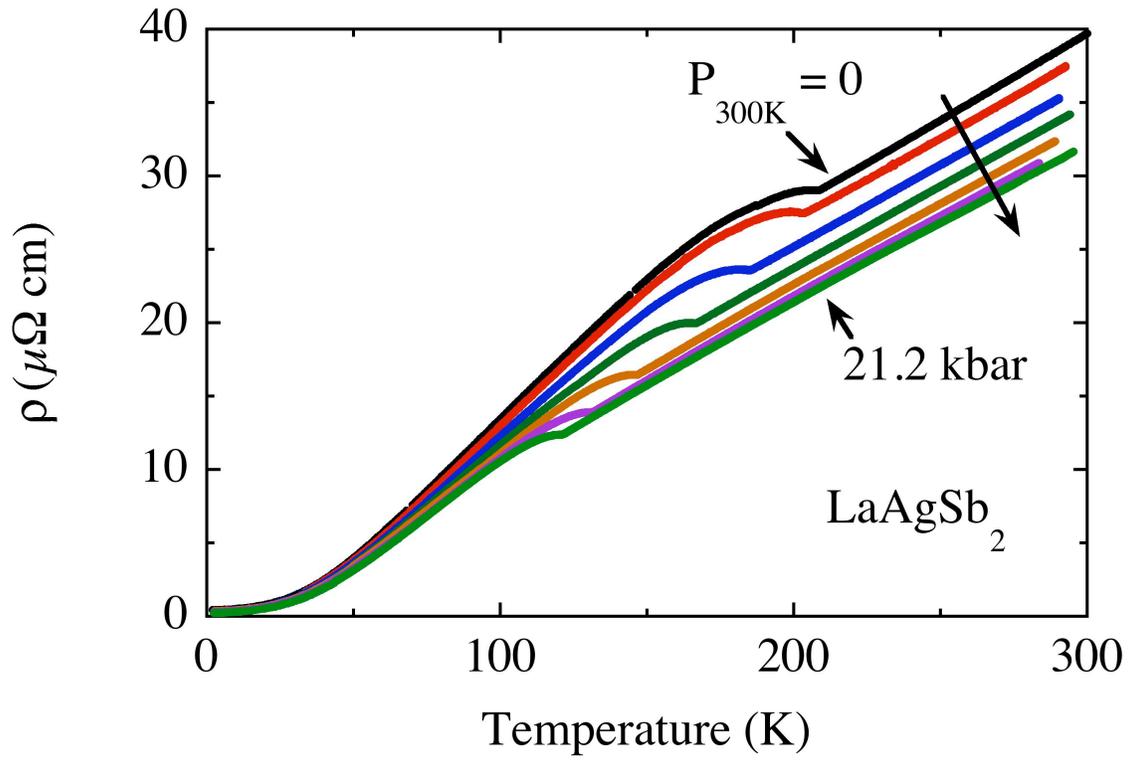

Figure 1

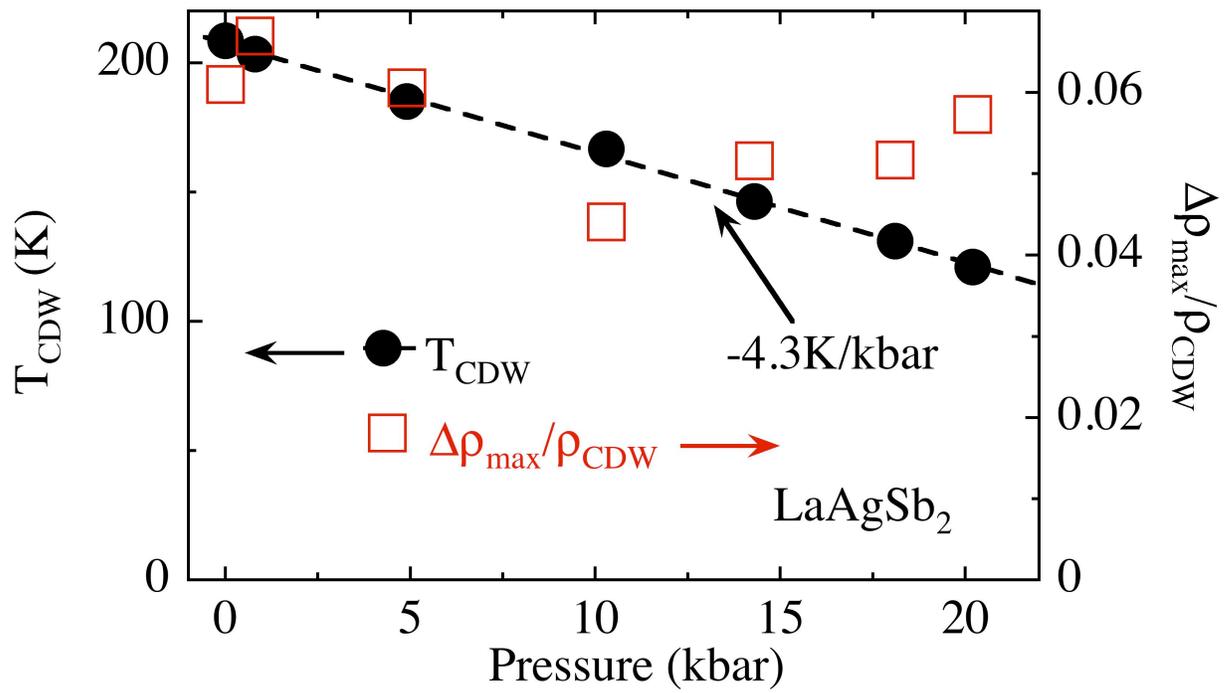

Figure 2

Torikachvili et. al.

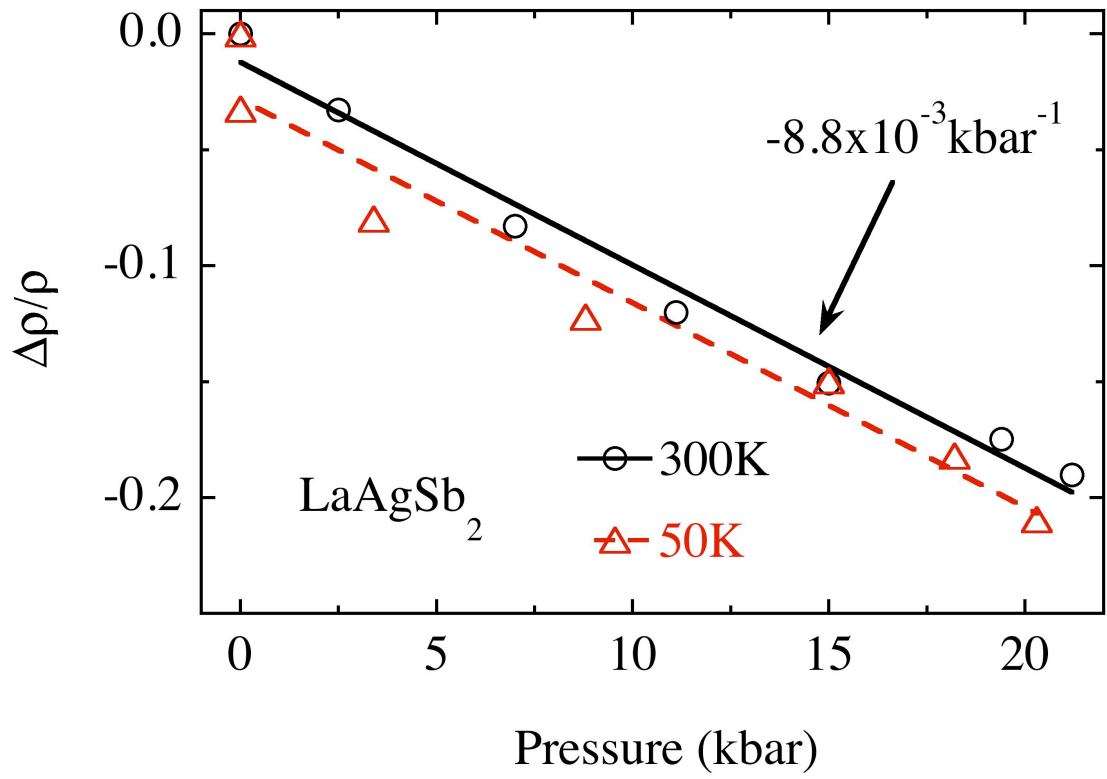

Figure 3

Torikachvili et. al.

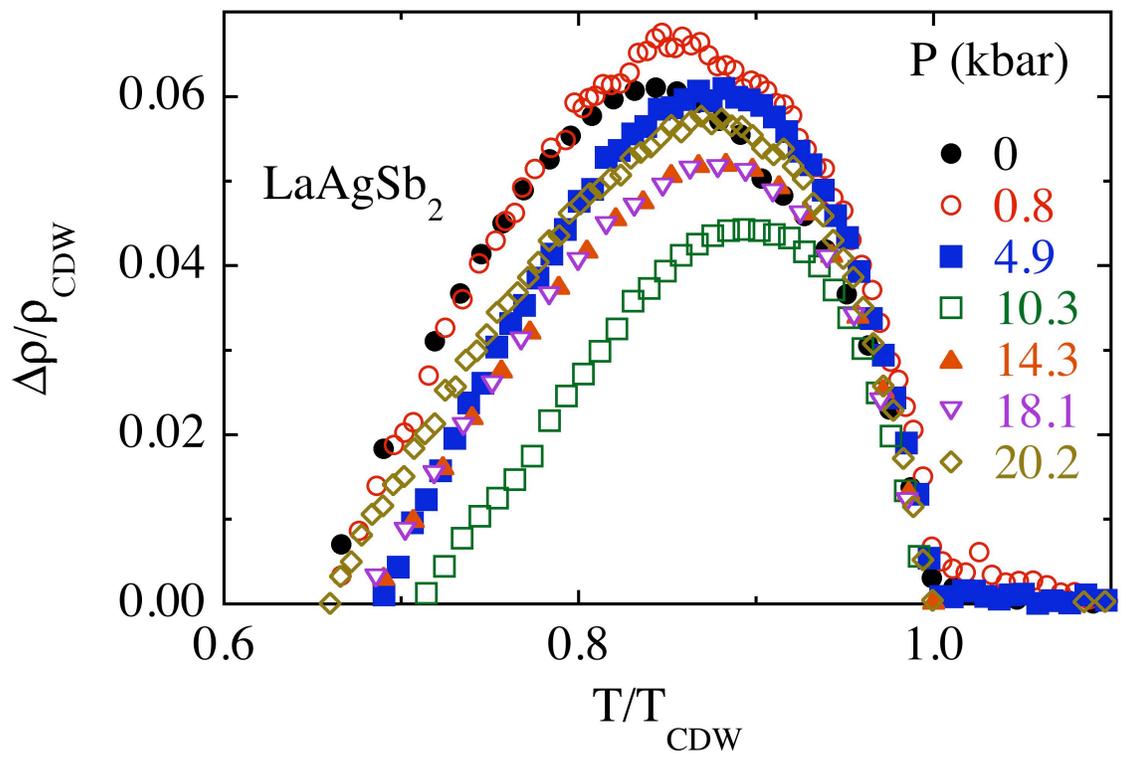

Figure 4

Torikachvili et al.

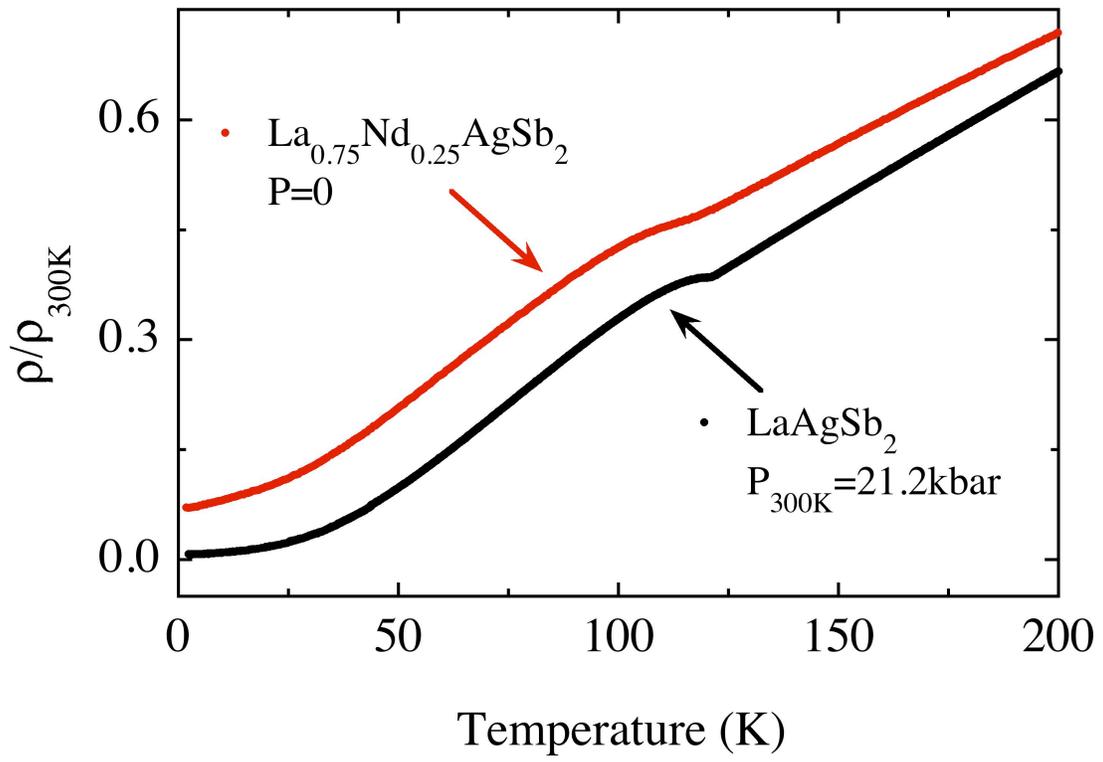

Figure 5

Torikachvili et al.

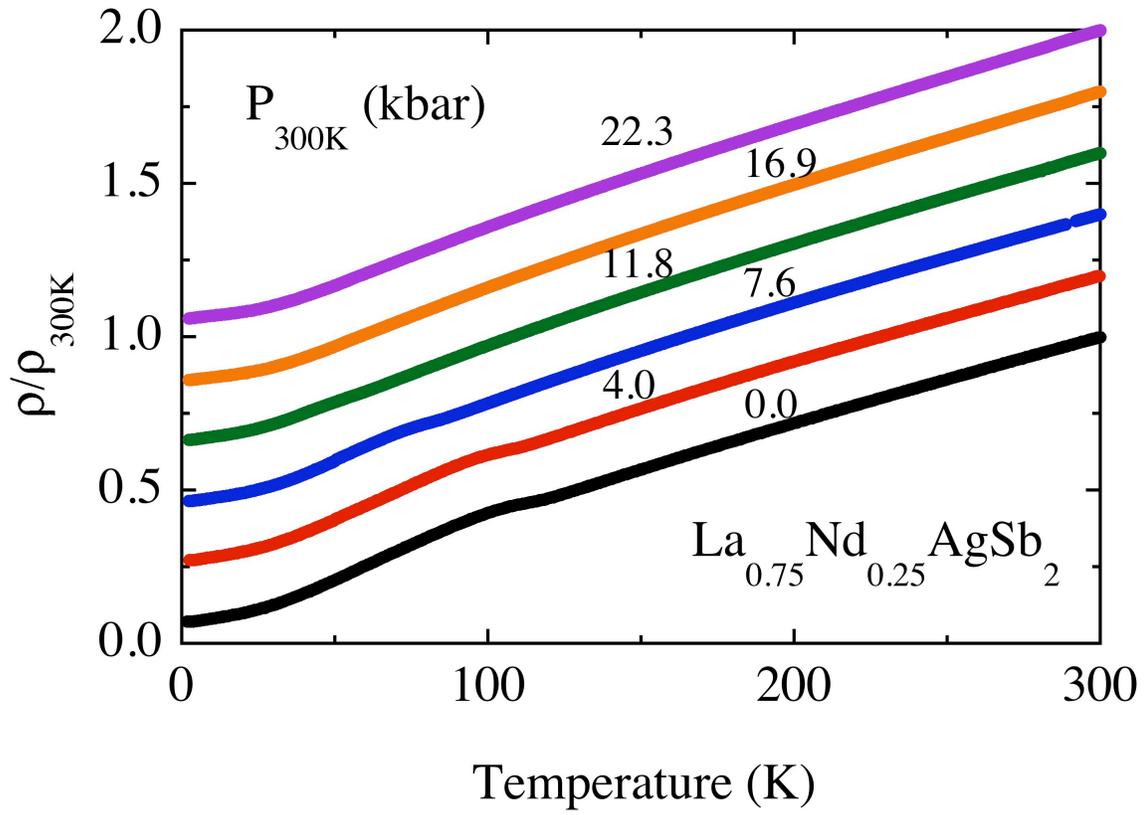

Figure 6

Torikachvili et al.

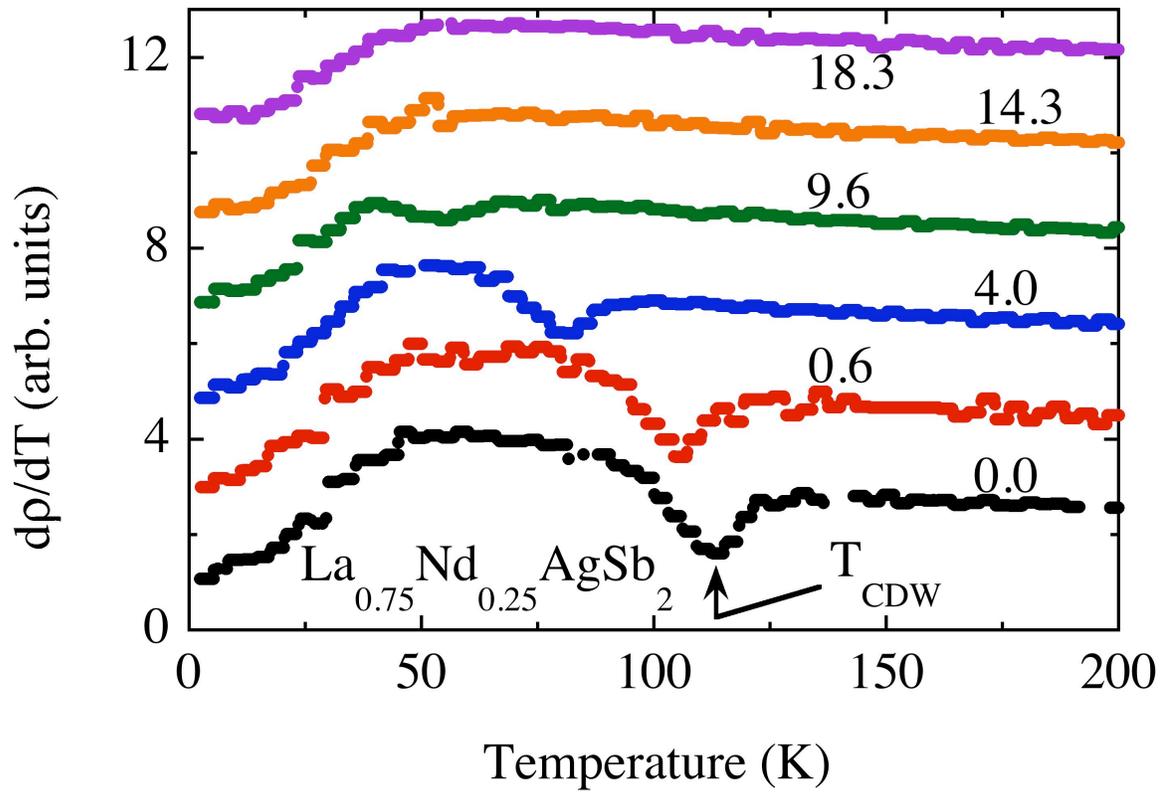

Figure 7

Torikachvili et al.

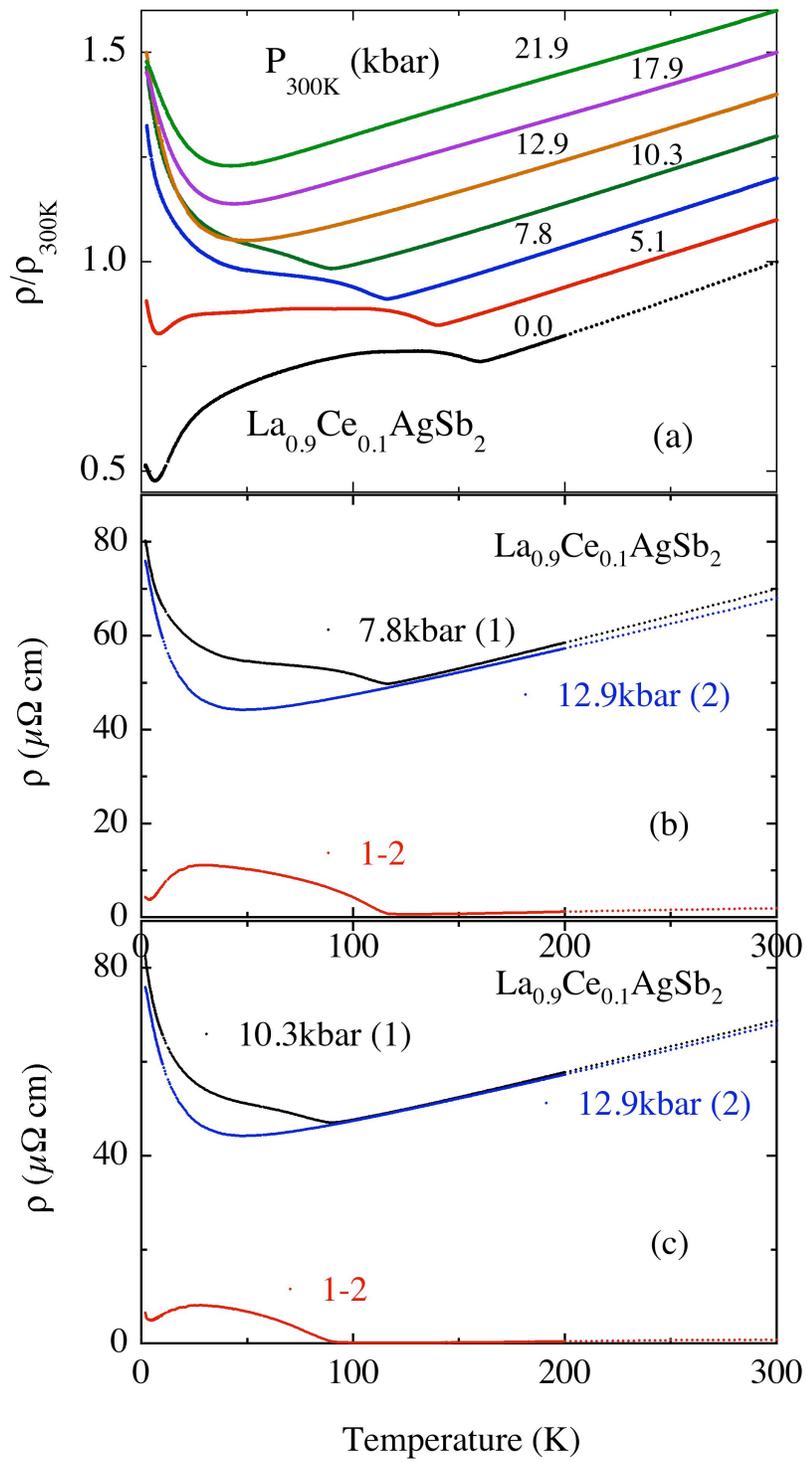

Figure 8
Torikachvili et al.

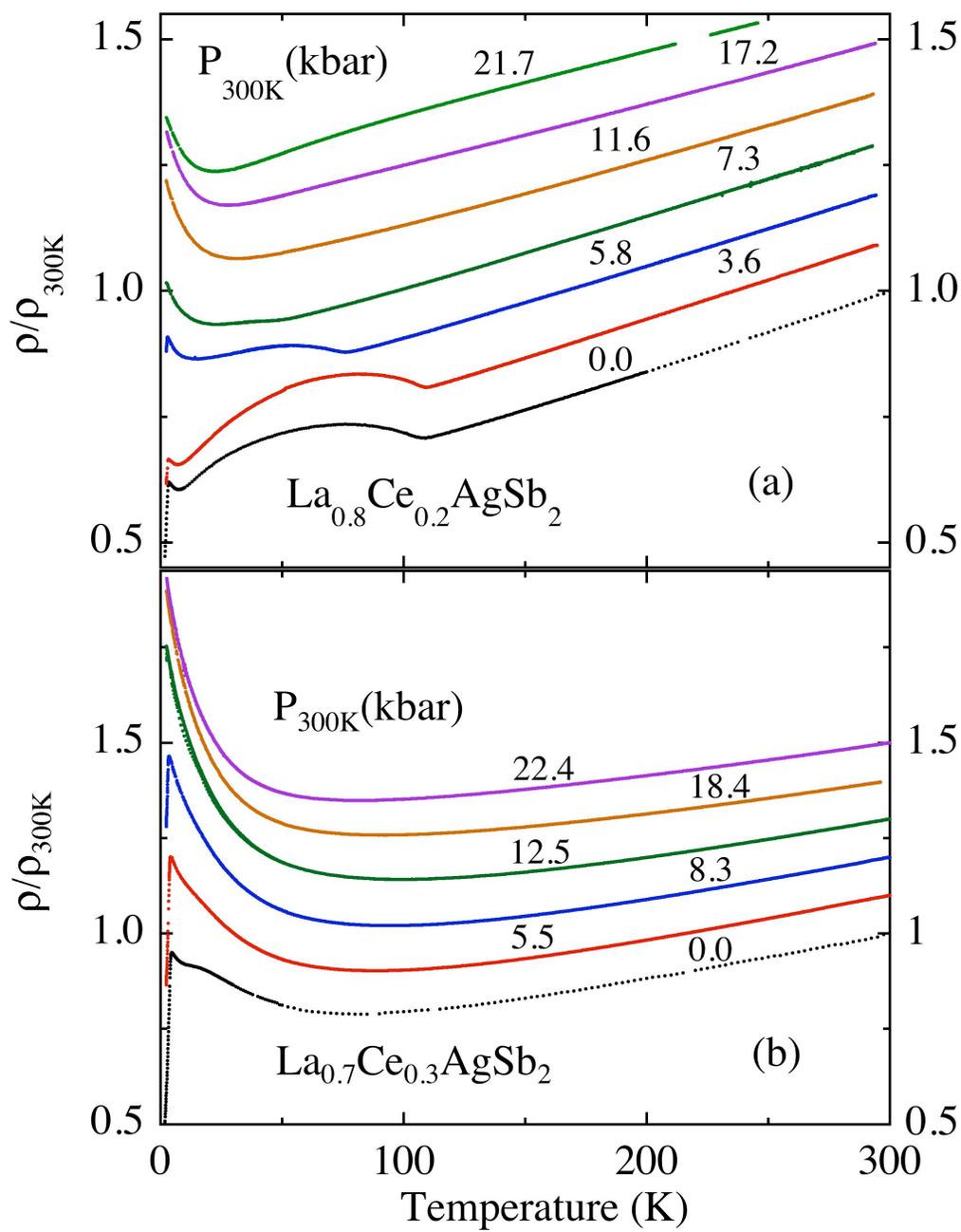

Figure 9

Torikachvili et al.

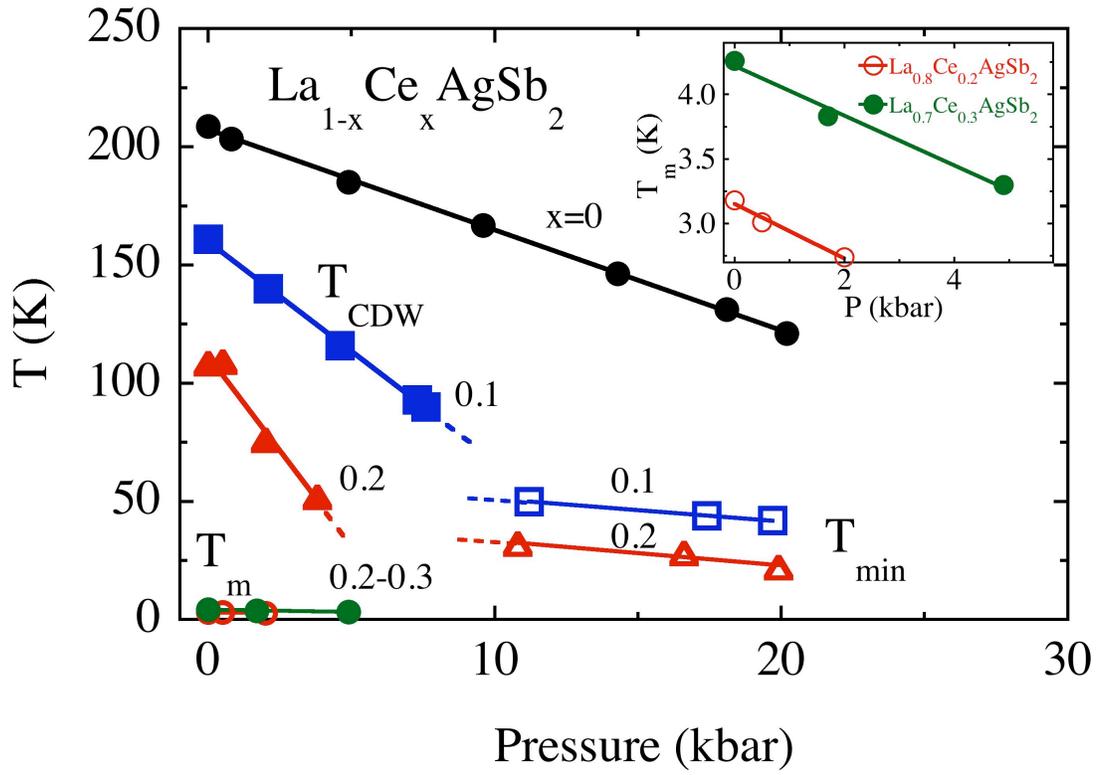

Figure 10

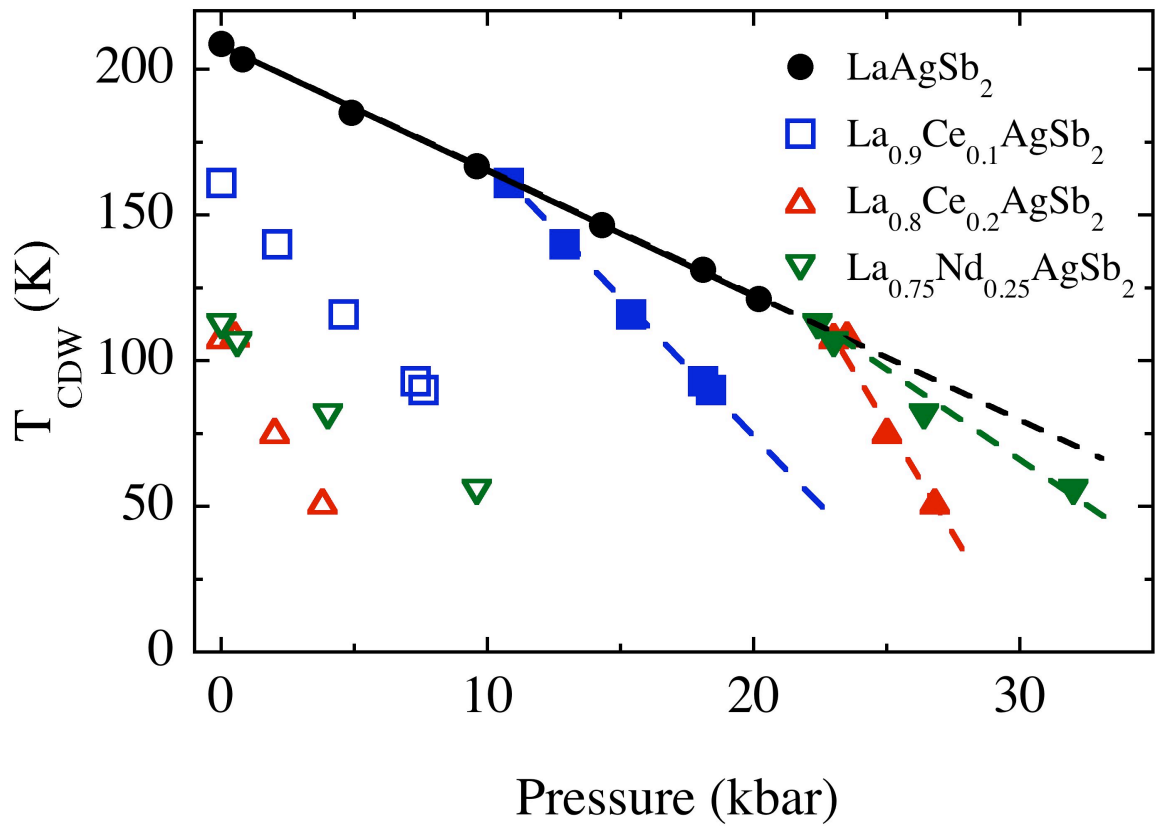

Figure 11